\documentclass{elsart}
\usepackage{graphics}
\usepackage{bm}

\begin{document}

\begin{frontmatter} 

\title{Neutrino magnetic moment effects in electron-capture 
measurements at GSI} 

\author{Avraham Gal}
%\ead{avragal@vms.huji.ac.il}
\address{Racah Institute of Physics, The Hebrew University, Jerusalem 
91904, Israel} 

\date{\today} 

\begin{abstract} 
Oscillatory behavior of electron capture rates in the two-body decay 
$D \to R + \nu$ of hydrogen-like ion into recoil ion plus undetected 
neutrino $\nu$, with a period of approximately 7~s, was reported in 
storage ring single-ion experiments at the GSI Laboratory, Darmstadt. 
Ivanov and Kienle [Phys. Rev. Lett. {\bf 103} (2009) 062502] have relegated 
this period to neutrino masses through neutrino mixing in the final state. 
New arguments are given here against this interpretation, while suggesting 
that these `GSI Oscillations' may be related to neutrino spin precession in 
the static magnetic field of the storage ring. This scenario requires a Dirac 
neutrino magnetic moment $\mu_{\nu}$ six times lower than the Borexino 
solar neutrino upper limit of $0.54 \times 10^{-10} \mu_B$ 
[Phys. Rev. Lett. {\bf 101} (2008) 091302], and its consequences are 
briefly explored. 
\end{abstract} 


\end{frontmatter}

\section{Introduction} 
\label{sec:intro} 

Measurements of weak interaction decay of multiply ionized heavy ions 
coasting in the ion storage-cooler ring ESR at the GSI laboratory, 
since the first report in 1992 \cite{GBB92}, open up new vistas for 
dedicated studies of weak interactions. In particular, electron 
capture (EC) decay rates in hydrogen-like and helium-like $^{140}$Pr 
ions have been recently measured for the first time \cite{LBG07} by 
following the motion of the decay ions (D) and the recoil ions (R). 
The overall decay rates $\lambda_{\rm EC}$ of these two-body 
$^{140}{\rm Pr} \to {^{140}{\rm Ce}} + \nu$ EC decays, in which no neutrino 
$\nu$ is detected, are well understood within standard weak interaction 
calculations of the underlying $e^-p\to\nu_e n$ reaction \cite{PKB08,IFR07}. 
EC decay rates reported subsequently in H-like and He-like $^{142}$Pm ions 
\cite{WGL09} are consistent with these $^{140}$Pr EC decay rate analyses. 
However, a time-resolved decay spectroscopy applied subsequently to the 
two-body EC decay of H-like $^{140}$Pr and $^{142}$Pm single ions revealed 
an oscillatory behavior, or more specifically a time modulation of the 
two-body EC decay rate \cite{LBW08}: 
\begin{equation} 
\lambda_{\rm EC}(t)=\lambda_{\rm EC}[1+a_{\rm EC}\cos(\omega_{\rm EC} t + 
\phi_{\rm EC})], 
\label{eq:omega} 
\end{equation} 
with amplitude $a_{\rm EC} \approx 0.2$, and angular frequency 
$\omega^{\rm lab}_{\rm EC} \approx 0.89~{\rm s}^{-1}$ 
(period $T^{\rm lab}_{\rm EC} \approx 7.1$~s) in the 
laboratory system which is equivalent in the rest frame of the decay ion 
to a minute energy $\hbar \omega_{\rm EC} \approx 0.84 \times 10^{-15}$~eV. 
Subsequent experiments on EC decays of neutral atoms in solid environment 
have found no evidence for oscillations with periodicities of this order of 
magnitude \cite{VCD08,FBH08}. Thus, the oscillations observed in the GSI 
experiment could have their origin in some characteristics of the H-like ions, 
produced and isolated in the ESR, and in the electromagnetic fields specific 
to the ESR which are not operative in normal laboratory experiments. Indeed, 
it is suggested here that the `GSI Oscillations' could be due to the static 
magnetic field, perpendicular to the ESR, which stabilizes and navigates the 
motion of the ions in the ESR. 

Several works, by Kienle and collaborators, relegated the `GSI Oscillations' 
to interference between neutrino mass eigenstates that evolve coherently 
from the electron-neutrino $\nu_e$ \cite{IRK08,Fab08,KKi08,IKP08,IK09,IK10}. 
This idea apparently also motivated the GSI experiment \cite{LBW08}. 
Such interferences, according to these works, lead to oscillatory behavior 
given by Eq.~(\ref{eq:omega}) with angular frequency $\omega_{\nu_e}$ 
where, again in the decay-ion rest frame, 
\begin{equation}  
\hbar\omega_{\nu_e}=\frac{\Delta (m_{\nu}c^2)^2}{2M_Dc^2}\approx 
0.29 \times 10^{-15}~{\rm eV}. 
\label{eq:delE} 
\end{equation}  
Here, $\Delta (m_{\nu}c^2)^2 = (0.76 \pm 0.02) \times 10^{-4}$~eV$^2$ is 
a neutrino squared-mass difference extracted from solar $\nu$ plus KamLAND 
reactor $\bar{\nu}$ data \cite{SNO08} for the two mass-eigenstate neutrinos 
that almost exhaust the coupling to $\nu_e$, and $M_D\approx 130$~GeV/$c^2$ 
is the mass of the decay ion $^{140}{\rm Pr}^{58+}$. Although the value of 
$\hbar\omega_{\nu_e}$ on the r.h.s. of Eq.~(\ref{eq:delE}) is about three 
times smaller than the value of $\hbar \omega_{\rm EC}$ required to resolve 
the `GSI Oscillation' puzzle, getting down to this order of magnitude 
nevertheless presents a remarkable achievement if correct.{\footnote
{Eq.~(\ref{eq:delE}) was also obtained by Lipkin \cite{Lip10} assuming 
interference between two unspecified components of the initial state with 
different momenta and energies that can both decay into the same final state, 
an electron neutrino and a recoil ion with definite energy and momentum. 
This scenario was criticized by Peshkin \cite{Pes08}.}} 

Other authors \cite{Giu08,BLS08,KKL08,CGL09,Mer09,Fla09} have rejected any 
link between neutrino mass eigenstates and the EC decay rate oscillatory 
behavior reported by the GSI experiment \cite{LBW08}, the underlying argument 
is that since no neutrino is singled out, the EC decay rate sums incoherently 
over amplitudes related to neutrino mass eigenstates, whereas any oscillatory 
behavior requires interference between such amplitudes. To be more specific, 
if the time-dependent EC transition amplitude $A_{\nu_e}(i \to f;~t)$, 
from initial state $i$ (D injected at time $t=0$) to a final state $f$ 
(R plus a coherent combination of neutrino mass eigenstates at time $t$), 
is written in terms of transition amplitudes $A_{\nu_j}(i\to f;~t)$ that 
involve mass-eigenstate neutrinos $\nu_j$: 
\begin{equation} 
A_{\nu_e}(i\to f;~t) = \sum_jU_{ej}A_{\nu_j}(i\to f;~t), 
\label{eq:Ae} 
\end{equation} 
where $U_{e j}$ are mixing elements of the $3 \times 3$ 
(assumed unitary) matrix $U$ 
\begin{equation} 
|\nu_\alpha\rangle = \sum_{j=1}^3 U_{\alpha j}^* \, |\nu_j\rangle~~~~
(\alpha = e, \mu, \tau) 
\label{eq:U} 
\end{equation} 
between the emitted $\nu_e$ and mass-eigenstate neutrinos 
$\nu_j$ \cite{ADA08}, then the associated probability is given by summing 
incoherently on $j=1,2,3$: 
\begin{equation} 
{\cal P}_{\nu_e}(i\to f;~t)=\sum_{j}|U_{ej}|^2|A_{\nu_j}(i\to f;~t)|^2
\approx |A_{\nu}(i\to f;~t)|^2, 
\label{eq:incoherent} 
\end{equation}  
where the dependence of the absolute-squared terms 
$|A_{\nu_j}(i\to f;~t)|^2$ on the species $\nu_j$ was neglected.{\footnote
{This neglect does not hold for interference terms $A_{\nu_j}A^*_{\nu_{j'}}$, 
$j \neq j'$, which give rise to oscillatory behavior, as discussed in 
Sect.~\ref{sec:osc}.}} 
According to Eq.~(\ref{eq:incoherent}), the probability ${\cal P}_{\nu_e}(i\to 
f;~t)$ for the two-body EC decay to occur is what standard weak interaction 
theory yields for a massless electron neutrino, regardless of its coupling to 
mass-eigenstate neutrinos. This holds true also for the total EC decay rate 
which is obtained by time differentiation of ${\cal P}_{\nu_e}(i\to f;~t)$ 
plus integration over phase space and which is found identical with the time 
independent decay rate $\lambda_{\rm EC}$ derived ignoring neutrino mixing. 

From the above discussion one notes that incoherence in terms of neutrino 
mass eigenstates rules out expressing the probability ${\cal P}_{\nu_e}$ 
as a squared absolute value of the amplitude $A_{\nu_e}$: 
\begin{equation} 
{\cal P}_{\nu_e}(i\to f;~t) \neq |A_{\nu_e}(i\to f;~t)|^2. 
\label{eq:Pe} 
\end{equation} 
It is instructive to ask whether incoherence shows up also in the flavor 
basis, since for times of order seconds which are appropriate to the `GSI 
Oscillations' the coherence implied by Eq.~(\ref{eq:Ae}) is still in effect 
and the flavor basis is of physical significance \cite{Mer09}. 
To this end I project Eq.~(\ref{eq:Ae}) onto flavor $\beta$: 
\begin{equation} 
A_{\nu_e \to \nu_{\beta}}(i\to f;~t) = \sum_jU_{ej}A_{\nu_j}(i\to f;~t)
U^*_{\beta j},  ~~~~(\beta = e, \mu, \tau) 
\label{eq:Abeta} 
\end{equation} 
in close analogy with the discussion of neutrino flavor oscillations 
in dedicated oscillation experiments (Eq.~(13.4) in Ref.~\cite{ADA08}). 
The summed probability to have any of these three flavors appear in the 
final state, without specifying which one, is obtained by squaring 
$|A_{\nu_e \to \nu_{\beta}}(i\to f;~t)|$ and summing over $\beta$: 
\begin{equation} 
\sum_{\beta}|\sum_j U_{ej}A_{\nu_j}(i\to f;~t)U^*_{\beta j}|^2 = 
\sum_{j}|U_{ej}|^2|A_{\nu_j}(i\to f;~t)|^2, 
\label{eq:interf} 
\end{equation} 
where the assumed unitarity of the mixing matrix $U$, 
$\sum_{\beta}U^*_{\beta j}U_{\beta j'}=\delta_{jj'}$, was instrumental in 
eliminating the interference terms, leading to an incoherent sum identical 
with ${\cal P}_{\nu_e}(i\to f;~t)$ of Eq.~(\ref{eq:incoherent}). 

The purpose of the present paper is twofold. First, to show that even if the 
arguments given above against coherence are disregarded, and one chooses 
to evaluate $|A_{\nu_e}(i\to f;~t)|^2$ as was done by Ivanov and Kienle in 
Ref.~\cite{IK09} contradicting Eq.~(\ref{eq:Pe}) here, the resulting 
oscillation period would be many orders of magnitude shorter than required 
to explain the `GSI Oscillations', and hence unobservable. More specifically, 
it is shown in Sect.~\ref{sec:osc} that the energy scale resulting by 
following the methodology of Ref.~\cite{IK09} is given by 
\begin{equation} 
\hbar\Omega_{\nu_e}=\frac{\Delta (m_{\nu}c^2)^2}{2E_{\nu}}\approx 0.95 
\times 10^{-11}~{\rm eV}, 
\label{eq:DelE} 
\end{equation} 
where $E_{\nu}\approx 4$~MeV is a representative value for neutrino 
energy in the H-like $^{140}{\rm Pr}\to{^{140}{\rm Ce}}+\nu_e$ and 
$^{142}{\rm Pm}\to {^{142}{\rm Nd}}+\nu_e$ EC decays \cite{LBW08}. 
The energy $\hbar\Omega_{\nu_e}$ is larger by over four orders of 
magnitude than $\hbar\omega_{\rm EC}$ or $\hbar\omega_{\nu_e}$ given 
by Eq.~(\ref{eq:delE}), and so it would lead to modulation period 
shorter by over four orders of magnitude than the 7~s period reported 
by the GSI experiment. Given a time measurement resolution of order 
$0.5$~s \cite{LBW08}, the effect of such oscillatory behavior would 
average out to zero. 

The main purpose of the present paper, however, is to introduce a new 
energy scale $\hbar\omega_{\mu_{\nu}}$, essentially given by the product 
of the neutrino magnetic moment $\mu_{\nu}$ (or rather its upper limit) 
and the static magnetic field $B$ which is perpendicular to the ESR. 
It is argued in Sect.~\ref{sec:magfield} that precession of the neutrino 
spin in this magnetic field induces interferences that might lead 
to oscillations of the required period, namely that $\hbar\omega_{\mu_{\nu}}$ 
is commensurate with $\hbar\omega_{\rm EC}$. The arguments provided in 
Sect.~\ref{sec:magfield} are rather schematic and, judging by the various 
referee reports which helped to shape the final form of this published 
version, may appear controversial to many experts in the Neutrino community. 
Nevertheless, as stated by the last referee ``it will no doubt create further 
discussions and opposing views" that ``might help in reaching the required 
consensus."

\section{Interference and time modulation of two-body EC rates} 
\label{sec:osc} 
 
Here I show that a correct application of of the formalism followed by 
Ivanov and Kienle \cite{IK09}, accepting it for the sake of argument, 
leads to oscillations with angular frequency $\hbar\Omega_{\nu_e}$, 
Eq.~(\ref{eq:DelE}); not with angular frequency $\hbar\omega_{\nu_e}$, 
Eq.~(\ref{eq:delE}), as claimed in Ref.~\cite{IK09}. To this end, I use as 
closely as possible their specific time-dependent first-order perturbation 
theory amplitudes $A_{\nu_j}(i\to f;~t)$: 
\begin{equation} 
A_{\nu_j}(i\to f;~t) = -i\int^t_0\langle 
f(\vec{q}\,)\nu_j(\vec{k}_j)|H_{e\nu_j}(\tau)|i(\vec{0}\,)\rangle d\tau, 
\label{eq:Aej} 
\end{equation} 
with a weak-interaction Hamiltonian for the leptonic transition  
$e^-\to\nu_j$ given by 
\begin{equation}  
{\cal H}_{e\nu_j}(\tau)=\frac{G_F}{\sqrt{2}}V_{ud}\int{d^3x
[\bar{\psi}_n\gamma^{\lambda}(1 - g_A\gamma^5) \psi_p]
[\bar{\psi}_{\nu_j}\gamma_{\lambda}(1 - \gamma^5)\psi_{e^-}]}. 
\label{eq:Hej} 
\end{equation} 
Here, $x=(\tau,\vec{x}\,)$, $G_F$ is the Fermi constant, $V_{ud}$ is the CKM 
matrix element, $g_A$ is the axial coupling constant, and with $\psi_n(x)$, 
$\psi_p(x)$, $\psi_{\nu_j}(x)$ and $\psi_{e^-}(x)$ denoting neutron, proton, 
mass-eigenstate neutrino $\nu_j$ and electron field operators, respectively. 
EC decays occur at any time $\tau$ within [$0,t$], from time $t'=0$ of 
injection of D into the ESR to time $t'=t$ of order seconds and longer at 
which the EC decay rate is evaluated. In the single-ion GSI experiment 
\cite{LBW08} the heavy ions revolve in the ESR with a period of order 
$10^{-6}$~s and their motion is monitored nondestructively once per 
revolution. The decay is defined experimentally by the {\it correlated} 
disappearance of D and appearance of R, but the appearance in the frequency 
spectrum is delayed by times of order 1~s needed to cool R. The order of 
magnitude of the experimental time resolution is similar, about 0.5~s, 
as reflected in the time intervals used to exhibit the experimental decay 
rates ${\cal R}(t)$ in Figs.~3,4,5 of Ref.~\cite{LBW08}. The time-averaged 
decay rates determined in the ESR appear to agree with those measured 
elsewhere, e.g. for $^{142}$Pm \cite{VCD08}, and this consistency suggests 
that details of kinematics and motion of the heavy ions in the storage ring 
affect little the overall decay rates which are evaluated here in conventional 
time-dependent perturbation theory. Therefore, it is plausible to assume that 
the evolution of the final state in these single-ion EC measurements at GSI 
proceeds over times of order 1~s which is used here as a working hypothesis. 

To obtain the time dependence of the amplitude $A_{\nu_j}(i \to f;~t)$ 
(similarly structured to Eq.~(6) of Ref.~\cite{IK09}), recall that the 
time dependence of the integrand in Eq.~(\ref{eq:Aej}) is given by 
$\exp({\rm i}\Delta_j\tau)$ where{\footnote{From here on $\hbar=c=1$ units 
are almost exclusively used.}} 
\begin{equation} 
\Delta_j(\vec{q}\,) = E_R(-\vec{q}\,) + E_j(\vec{q}\,) - M_D 
\label{eq:DeltaE} 
\end{equation} 
with 
\begin{equation} 
E_R=\sqrt{M_R^2+(-\vec{q}\,)^2}, \,\,\,\,\,\,
E_j=\sqrt{m_j^2+\vec{q}\,^2} 
\label{eq:E_Rj} 
\end{equation} 
for the recoil ion and neutrino $\nu_j$ energies, respectively, 
in the decay-ion rest frame. Integrating on this time dependence 
results in a standard time-dependent perturbation-theory energy-time 
dependence \cite{Baym69} 
\begin{equation} 
A_{\nu_j}(i \to f;~t) \sim \frac{1 - \exp({\rm i}\Delta_j t)}{\Delta_j}. 
\label{eq:TD1} 
\end{equation} 
The EC decay rate ${\cal R}_{\nu_e}(i\to f;~t)$ is obtained 
from the probability ${\cal P}_{\nu_e}(i\to f;~t)$, Eq.~(\ref{eq:incoherent}), 
by differentiating: ${\cal R}={\partial}_t{\cal P}$. Using Eq.~(\ref{eq:TD1}) 
for the time dependence of $A_{\nu_j}(i \to f;~t)$, one gets a nonoscillatory 
contribution to ${\cal R}_{\nu_e}$: 
\begin{equation} 
{\cal R}_{\nu_j} = \frac{d}{dt}|A_{\nu_j}(i \to f;~t)|^2 \sim 
\frac{2\sin (\Delta_j t)}{\Delta_j} \to 2\pi\delta(\Delta_j), 
\label{eq:TD2} 
\end{equation} 
where the last step requires a sufficiently long time $t$. 
The properly normalized contribution of these terms to 
${\cal R}_{\nu_e}(i\to f;~t)$ is given by 
\begin{equation} 
\sum_j {\cal R}_{\nu_j} = \lambda_{\rm EC}\sum_j |U_{ej}|^2 \delta(\Delta_j) 
\approx \lambda_{\rm EC}\delta(\Delta),  
\label{eq:jj} 
\end{equation} 
where the dependence of $\delta(\Delta_j)$ on the species $j$ could be safely 
neglected. If $j' \neq j$ interference terms are considered, then their 
properly normalized contribution to ${\cal R}_{\nu_e}(i\to f;~t)$, again for 
sufficiently long times, is given by 
\begin{equation} 
\lambda_{\rm EC}\sum_{j>j'} {\rm Re}(U_{ej}U_{ej'}^*) [\delta(\Delta_j)+
\delta(\Delta_{j'})] \cos[(\Delta_j - \Delta_{j'})t]. 
\label{eq:ivanov} 
\end{equation} 
The Dirac $\delta$ functions in Eqs.~(\ref{eq:jj}) and (\ref{eq:ivanov}) take 
care of energy conservation and have to be integrated upon, instead of the 
more customary integration on the implied c.m. momentum $\vec q$ to obtain 
the EC decay rate. It is straightforward to integrate over $\Delta$ for the 
nonoscillatory terms which then yield as expected the rate $\lambda_{\rm EC}$ 
in Eq.~(\ref{eq:jj}). For the oscillatory terms it is impossible to satisfy 
both $\delta(\Delta_j)$ and $\delta(\Delta_{j'})$ {\it simultaneously} in 
Eq.~(\ref{eq:ivanov}), meaning that the phase $(\Delta_j - \Delta_{j'})t$ is 
once evaluated under the constraint $\Delta_j(\vec{q}\,)=0$ and once under 
the constraint $\Delta_{j'}(\vec{q}\,)=0$. On each occasion, using a generic 
notation $k$ for the momentum implied by each one of the $\delta$ functions, 
one obtains to an excellent approximation 
\begin{equation} 
\Delta_j(k)-\Delta_{j'}(k)= E_j(k)-E_{j'}(k) = \hbar \Omega_{jj'}, 
\label{eq:correct} 
\end{equation} 
where $\Omega_{jj'}$ is related to $\Omega_{\nu_e}$ of 
Eq.~(\ref{eq:DelE}): 
\begin{equation} 
\hbar \Omega_{jj'} = \frac{m_j^2-m_{j'}^2}{2E_{\nu}} \approx \hbar 
\Omega_{\nu_e}. 
\label{eq:Omegajj'} 
\end{equation} 
Ivanov and Kienle \cite{IK09} overlooked this subtlety by using in 
Eq.~(\ref{eq:correct}) simultaneously {\it on energy shell} momentum values 
$k_j$ and $k_{j'}$ implied by $\delta(\Delta_j)$ and $\delta(\Delta_{j'})$ 
respectively, and replacing $\Delta_j - \Delta_{j'}$  in the oscillatory 
terms of Eq.~(\ref{eq:ivanov}) by $E_j(k_j)-E_{j'}(k_{j'})\approx \hbar 
\omega_{\nu_e}$, Eq.~(\ref{eq:delE}). A similar error was made by 
Kleinert and Kienle when evaluating Eq.~(54) in Ref.~\cite{KKi08}. 

The requirement of {\it sufficiently long times} for Eq.~(\ref{eq:ivanov}) 
to hold translates in the present case to requiring $t\gg\Omega_{\nu_e}^{-1}
\sim 7\times 10^{-5}$~s, which is comfortably satisfied given the experimental 
time resolution scale of $\sim 0.5$~s \cite{LBW08}. Furthermore, as already 
discussed in Sect.~\ref{sec:intro}, oscillations with periodicities of order 
$10^{-4}$~s would average out to zero in the GSI experiments, even if 
conceptually allowed.

\section{Magnetic field effects} 
\label{sec:magfield}

The preceding discussion ignored a possible role of the electromagnetic 
fields surrounding the ESR for guidance and stabilization of the 
heavy-ion motion. The nuclei $^{140}$Pr and $^{142}$Pm in the GSI 
experiment \cite{LBW08} have spin-parity $I_i^{\pi}=1^+$, and the 
electron-nucleus hyperfine interaction in the decay ion forms a doublet 
of levels $F_i^{\pi}=({\frac{1}{2}}^+,{\frac{3}{2}}^+)$, 
the `sterile' ${\frac{3}{2}}^+$ level lying about 1 eV above the `active' 
${\frac{1}{2}}^+$ g.s. from which EC occurs to a $F_f=\frac{1}{2}$ final 
state of a fully ionized recoil ion with spin-parity $I_f^{\pi}=0^+$ plus 
a left-handed neutrino of spin $\frac{1}{2}$.{\footnote{The subscript $f$ 
in this section relates to both the recoil ion and the neutrino.}} The 
lifetime of the $F_i^{\pi}={\frac{3}{2}}^+$ excited level is of order 
$10^{-2}$~s, so that it de-excites sufficiently rapidly to the 
$F_i^{\pi}={\frac{1}{2}}^+$ g.s. \cite{LBG07,IFR07}. 
Periodic excitations of this `sterile' state cannot explain the reported 
time dependence and intensity pattern \cite{WSI09}. 
The static magnetic field which is perpendicular to the ESR, $B=1.19$~T 
for $^{140}{\rm Pr}$ \cite{Fas09private}, gives rise to precession of the 
$F_i^{\pi}={\frac{1}{2}}^+$ initial-state spin with angular frequency 
$\omega_i$ of order $\hbar\omega_i \sim {\mu_B}B\approx 0.7\times 10^{-4}$~eV 
\cite{FIK09}, where $\mu_B$ is the Bohr magneton. The corresponding time scale 
of order $10^{-11}$~s is substantially shorter than even the ESR revolution 
period $t_{\rm revol} \approx 0.5\times 10^{-6}$~s, so any oscillation arising 
from this initial-state precession would average out to zero over 1 cm of the 
approximately 100 m long circumference. A nonstatic magnetic field could lead 
through its high harmonics to oscillations with the desired frequency between 
the magnetic substates of the $F_i^{\pi}={\frac{1}{2}}^+$ g.s. \cite{Pav10}, 
but the modulation amplitude $a_{\rm EC}$ expected for such harmonics is 
substantially below a $1\%$ level, and hence negligible. 
Furthermore, the associated mixing between the two hyperfine levels 
$F_i^{\pi}=({\frac{1}{2}}^+,{\frac{3}{2}}^+)$ is negligible. In conclusion, 
no initial-state coherence effects are expected from internal or external 
electromagnetic fields in the GSI experiment. 

In the final configuration, interferences may arise from the precession of 
the neutrino spin in the static magnetic field of the ESR.{\footnote{In 
disagreement with Merle's recent claim ``a splitting in the final state cannot 
explain the GSI oscillations" \cite{Mer09} which ignored electromagnetic 
effects.}} The corresponding angular frequency $\omega_{\mu_{\nu}}$ is given 
by $\hbar\omega_{\mu_{\nu}} = {\mu_{\nu}}\gamma B <0.5 \times 10^{-14}$~eV 
in the decay ion rest frame, due to the neutrino anomalous magnetic moment 
$\mu_{\nu}$ interacting with the static magnetic field $B$. 
Here, $\gamma=1.43$ is the Lorentz factor relating the rest frame 
to the laboratory frame, and $\mu_{\nu}< 0.54\times 10^{-10}\mu_B$ from the 
Borexino solar neutrino data \cite{Borexino08}. Below I show explicitly how 
the total EC rate gets time-modulated with angular frequency 
$\omega_{\mu_{\nu}}$. To agree with the reported GSI measurements, 
$\omega_{\mu_{\nu}}= \omega_{\rm EC}$, a value of the electron-neutrino 
magnetic moment $\mu_{\nu} \sim 0.9 \times 10^{-11} \mu_B$ is required which 
is six times lower than provided by the published Borexino solar neutrino 
upper limit \cite{Borexino08}.

\subsection{Interference due to a Dirac neutrino magnetic moment} 
\label{subsec:dirac}

For definiteness I first assume that neutrinos are Dirac fermions with only 
diagonal magnetic moments $\mu_{jk}=\mu_j \delta_{jk}$, and that these 
diagonal moments are the same for all three species: $\mu_j=\mu_{\nu}$. The 
emitted electron-neutrino is a left-handed lepton. The amplitude for producing 
it right-handed, namely with a positive helicity is negligible, of order 
$m_{\nu}/E_{\nu} < 10^{-7}$ and thus may be safely ignored. A static magnetic 
field perpendicular to the ESR flips the neutrino spin. Each of the 
mass-eigenstate components of the emitted neutrino will then precess, 
with amplitude $\cos (\omega_{\mu_{\nu}}\tau)$ for the depleted left-handed 
components and with amplitude ${\rm i}\sin (\omega_{\mu_{\nu}}\tau)$ for the 
spin-flip right-handed components \cite{FSh80}. Both are legitimate neutrino 
final states which are summed upon {\it incoherently}. 
The summed probability is of course time independent: 
$\cos^2(\omega_{\mu_{\nu}}\tau)+\sin^2 (\omega_{\mu_{\nu}}\tau)=1$. 
However, the magnetic field dipoles of the storage ring do not cover its full 
circumference, except for about $35\%$ of it \cite{Fas09private}. This results 
in interference between the decay amplitude $A^0_{\nu_j}$, for events with no 
magnetic interaction, and the decay amplitude $A^{\rm m}_{\nu_j}$ for events 
undergoing magnetic interaction with depleted left-handed components: 
\begin{equation} 
A^0_{\nu_j} \sim -{\rm i}\int^t_0 \exp({\rm i}\Delta_j\tau) d\tau, \,\,\,\, 
A^{\rm m}_{\nu_j} \sim -{\rm i}\int^t_0 \exp({\rm i}\Delta_j\tau) 
\cos (\omega_{\mu_{\nu}}\tau) d\tau,   
\label{eq:Aj} 
\end{equation} 
using the same normalization as in Eq.~(\ref{eq:TD1}) for any of the 
left-handed mass-eigenstate neutrinos. This expression for $A^{\rm m}_{\nu_j}$ 
represents physically the action of the magnetic field at time $\tau$ of 
the EC decay.{\footnote{See Ref.~\cite{0809.1213} for a different choice 
of $A^{\rm m}_{\nu_j}$ that yields, nevertheless, the same time modulation 
as given by Eq.~(\ref{eq:ratenu}) below.}} The related amplitude 
$A^{\rm R}_{\nu_j}$ for events undergoing magnetic interaction which have 
resulted in a right-handed neutrino is then given by: 
\begin{equation} 
A^{\rm R}_{\nu_j} \sim -{\rm i}\int^t_0 \exp({\rm i}\Delta_j\tau) 
{\rm i}\sin (\omega_{\mu_{\nu}}\tau) d\tau. 
\label{eq:AjR} 
\end{equation} 
Repeating the same steps in going from amplitudes $A_{\nu_j}$, 
Eq.~(\ref{eq:TD1}), to decay rates ${\cal R}_{\nu_j}$, Eq.~(\ref{eq:TD2}), 
and adopting the same normalization, the decay rates associated with each 
one of these three amplitudes are given by: 
\begin{equation} 
{\cal R}^0_{\nu_j} = \frac{d}{dt}|A^0_{\nu_j}|^2 \sim 2\pi\delta(\Delta_j), 
\label{eq:rate0j} 
\end{equation} 
\begin{equation} 
{\cal R}^{\rm m}_{\nu_j} = \frac{d}{dt}|A^{\rm m}_{\nu_j}|^2 
\sim \frac{\pi}{2}[\delta(\Delta_j+\omega_{\mu_{\nu}}) 
+ \delta(\Delta_j-\omega_{\mu_{\nu}})](1+\cos(2\omega_{\mu_{\nu}}t)), 
\label{eq:ratemj} 
\end{equation} 
\begin{equation} 
{\cal R}^{\rm R}_{\nu_j} = \frac{d}{dt}|A^{\rm R}_{\nu_j}|^2 
\sim \frac{\pi}{2}[\delta(\Delta_j+\omega_{\mu_{\nu}}) 
+ \delta(\Delta_j-\omega_{\mu_{\nu}})](1-\cos(2\omega_{\mu_{\nu}}t)). 
\label{eq:rateRj} 
\end{equation} 
Note that although the two latter expressions for rates associated with the 
magnetic interaction are time dependent, their sum is time independent as 
expected from summing incoherently over the two separate helicities. The only 
time dependence in this schematic model arises from interference of the two 
amplitudes $A^0_{\nu_j}$ and $A^{\rm m}_{\nu_j}$ for a left-handed neutrino. 
Incorporating this interference, the total EC decay rate corresponding to 
$\nu_j$ is given by
\begin{eqnarray} 
{\cal R}_{\nu_j}=\frac{d}{dt}(|a_0 A^0_{\nu_j}+a_{\rm m} A^{\rm m}_{\nu_j}|^2
+|A^{\rm R}_{\nu_j}|^2) \nonumber \\ 
\sim |a_0|^2 2\pi\delta(\Delta_j) + |a_{\rm m}|^2 \pi 
[\delta(\Delta_j+\omega_{\mu_{\nu}}) + \delta(\Delta_j-\omega_{\mu_{\nu}})] 
\nonumber \\ 
+ 2{\rm Re}(a_0 a^*_{\rm m}) \frac{\pi}{2}
[\delta(\Delta_j+\omega_{\mu_{\nu}}) + 2\delta(\Delta_j) 
+ \delta(\Delta_j-\omega_{\mu_{\nu}})] \cos (\omega_{\mu_{\nu}}t) 
\nonumber \\ 
+ 2{\rm Im}(a_0 a^*_{\rm m}) \frac{\pi}{2}
[\delta(\Delta_j+\omega_{\mu_{\nu}}) - \delta(\Delta_j-\omega_{\mu_{\nu}})] 
\sin (\omega_{\mu_{\nu}}t), 
\label{eq:ratej} 
\end{eqnarray} 
where $|a_{\rm m}|^2 \sim 0.35$ and $|a_0|^2 \sim 0.65$, with unknown relative 
phase between the probability amplitudes $a_{\rm m}$ and $a_0$ for undergoing 
or not undergoing magnetic interaction, respectively. 
Working out the complete normalization of this expression, the final rate 
expression is given by 
\begin{equation} 
{\cal R}_{\nu_e} = \lambda_{\rm EC}[1+2{\rm Re}(a_0 a^*_{\rm m}) 
\cos (\omega_{\mu_{\nu}}t)],  
\label{eq:ratenu} 
\end{equation} 
showing explicitly a time modulation of the kind Eq.~(\ref{eq:omega}) reported 
by the GSI experiment \cite{LBW08}. It is beyond the present schematic model 
to explain the magnitude of the modulation amplitude $a_{\rm EC}$ and the 
phase shift $\phi_{\rm EC}$, except that $|a_{\rm EC}|<1$. In particular, 
a more realistic calculation is required in order to study effects of 
departures from the idealized kinematics implicitly considered above by which 
{\it both} the recoil ion and the neutrino go forward with respect to the 
decay-ion instantaneous laboratory forward direction. Whereas this is an 
excellent approximation for the recoil-ion motion, it is less so for the 
neutrino.{\footnote{I owe this observation to Eli Friedman.}} Nevertheless, 
for a rest-frame isotropic distribution, it is estimated that neutrino forward 
angles in the laboratory dominate over backward angles by more than a factor 
five. 

For distinct diagonal Dirac-neutrino magnetic moments, Eq.~(\ref{eq:ratenu}) 
gets generalized to 
\begin{equation} 
{\cal R}_{\nu_e} = \lambda_{\rm EC}[1+2{\rm Re}(a_0 a^*_{\rm m}) 
\sum_j |U_{ej}|^2 \cos (\omega_{\mu_j}t)],  
\label{eq:ratenuj} 
\end{equation} 
resulting in a more involved pattern of modulation. Finally, for vanishing 
diagonal magnetic moments, and nonzero values of transition magnetic moments, 
the discussion proceeds identically to that for Majorana neutrinos in the next 
subsection.

\subsection{Majorana neutrino magnetic moments} 
\label{subsec:majorana} 

Majorana neutrinos can have no diagonal electromagnetic moments, but are 
allowed to have nonzero {\it transition} moments connecting different 
mass-eigenstate neutrinos, or different flavor neutrinos. A static magnetic 
field perpendicular to the storage ring will induce spin-flavor precession 
\cite{SVa81}. However, the magnetic interaction effect is masked in this case 
by neutrino mass differences, such that the amplitudes 
$\cos (\omega_{\mu_{\nu}}\tau)$ and $\sin (\omega_{\mu_{\nu}}\tau)$ in 
Eqs.~(\ref{eq:Aj}) and (\ref{eq:AjR}) are replaced, to leading order in 
$\omega_{\mu_{\nu}}/\Omega_{\nu_e} << 1$, by 
\begin{equation} 
\cos (\omega_{\mu_{\nu}}\tau) \rightarrow \exp(-{\rm i}\Omega_{jj'}\tau),
\,\,\,\,\, 
\sin (\omega_{\mu_{\nu}}\tau) \rightarrow 
\frac{\omega_{\mu_{jj'}}}{\Omega_{jj'}}\sin (\Omega_{jj'}\tau),  
\label{eq:transition} 
\end{equation} 
where $\hbar \omega_{\mu_{jj'}} = \mu_{jj'}\gamma B$, and $\Omega_{jj'}$ is 
defined by Eq.~(\ref{eq:correct}). 
The period of any oscillation that might be induced by these amplitudes 
is of order $\Omega_{\nu_e}^{-1}\sim 7\times 10^{-5}$~s which is several 
orders of magnitude shorter than the time resolution scale of $\sim 0.5$~s 
in the GSI experiment \cite{LBW08}. Therefore, such oscillations will 
completely average out to zero over realistic detection periods.

\section{Discussion and summary} 
\label{sec:sum} 

In this work I have discussed several interference scenarios which might 
be of relevance to the issue of `GSI Oscillations'. It was reaffirmed that 
interference terms between different propagating mass-eigenstate neutrino 
amplitudes in two-body EC reactions on nuclei do not arise when no particular 
neutrino is singled out. A cancellation of such interference terms occurs also 
within a flavor oriented discussion, requiring however that the neutrino 
mass-flavor mixing matrix $U$ is unitary. Interference terms of this kind 
arise and give rise to oscillatory behavior of the EC decay rate, if and 
{\it only} if a particular neutrino flavor is singled out. It was shown here 
and in Ref.~\cite{0809.1213} that the relevant period of oscillations is 
$T\sim 4\pi E_{\nu}/{\Delta (m_{\nu}^2)}$ which for 
$E_{\nu}\approx 4$~MeV as in the GSI experiments \cite{LBW08}, 
and for $\Delta (m_{\nu}^2) \approx 0.76 \times 10^{-4}$~eV$^2$ \cite{SNO08}, 
assumes the value $T\sim 4.4\times 10^{-4}$~s, shorter by over four orders 
of magnitude than the period reported in these experiments. The oscillation 
period cited here is in full agreement with the oscillation length tested 
in dedicated neutrino oscillation experiments,{\footnote{Detailed expressions 
are given in Eqs.~(20,21,22) of Ref.~\cite{0809.1213} where a more rigorous 
wave-packet treatment would be required to justify the transition from $t$ to 
$L$ \cite{AS09}.}} provided the time $t$ is identified with $L/c$ where $L$ 
is the distance traversed by the neutrino between its source and the detector.
In particular, besides the $\Delta (m_{\nu}^2)$ neutrino input, it depends 
on the neutrino energy $E_{\nu}$, not on the mass $M_D$ of the decay ion. 

On the positive side, I have proposed a possible explanation of 
the `GSI Oscillations' puzzle connected with the magnetic field that guides 
the heavy-ion motion in the ESR, requiring a Dirac neutrino magnetic moment 
$\mu_{\nu}$ about six times lower than the laboratory upper limit value 
from the Borexino Collaboration \cite{Borexino08}. The underlying mechanism 
is the interference between EC decay amplitudes not affected by the static 
magnetic field of the ESR and EC decay amplitudes affected by this field 
which induces spin precession of the emitted neutrino. Each of the outgoing 
neutrinos, provided it is left-handed, has two {\it indistinguishable} `paths' 
to go through the ESR once it is produced in the EC decay: one is to encounter 
the static perpendicular magnetic field of the ESR, the other is to miss it. 
This is precisely like in the two-slit experiment. Interference is unavoidable 
then and is recorded by the motion of the entangled recoil ion in the ESR long 
after the neutrino has fled away. 

The spin-precession interference mechanism does not work for Majorana 
neutrinos that may have only {\it transition} magnetic moments. The resulting 
spin-flavor precession is suppressed by neutrino mass differences, and it 
becomes impossible to relate then the GSI Oscillations puzzle to magnetic 
effects. It is not yet resolved experimentally whether neutrinos are Dirac 
or Majorana fermions, although the theoretical bias rests with Majorana 
fermions, in which case the present paper accomplished nothing towards 
providing a credible explanation of this puzzle. 

For experimental verification, note that the time-modulation period 
$T^{\rm lab}_{\rm EC}$ is inversely proportional to $B$, so the effect 
proposed here may be checked by varying $B$, for example by varying 
$\beta=v/c$ for the coasting decay ions. For a fixed value of $\beta$, 
$B$ depends on the charge-to-mass ratio of the decay ion which varies 
only to a few percent with the decay-ion mass $M_D$. Finally, the proposed 
effect is unique to two-body EC reactions, since three-body weak decays do 
not constrain the neutrino direction of motion with respect to the fixed 
direction of $\vec B$. Indeed, preliminary data on the three-body $\beta^+$ 
decay of $^{142}$Pm indicate no time modulation of the $\beta^+$ decay rate, 
limiting its modulation amplitude to $a_{\beta^+} < 0.03(3)$ \cite{Kienle09}.

\section*{Acknowledgement} Critical comments by Thomas Faestermann, Eli 
Friedman and Koichi Yazaki, and stimulating discussions with Paul Kienle and 
Harry Lipkin, are greatly appreciated. Special thanks are due to Wolfram Weise 
and members of the T39 Group for the kind hospitality during a three-month 
visit in 2009 to the Technische Universit\"{a}t M\"{u}nchen (TUM), where 
Sect.~\ref{sec:magfield} was conceived. The support by the DFG Cluster of 
Excellence `Origin and Structure of the Universe' at TUM is gratefully 
acknowledged.

\end{document}